\documentclass[referee]{raa}           
\usepackage{graphicx,times}
\usepackage{natbib}
\usepackage{amssymb,amsmath}
\bibpunct{(}{)}{;}{a}{}{,}

\usepackage{caption}
\usepackage[papersize={9in,11in}]{geometry}
\usepackage{hyperref}
\hypersetup{pdftitle = The title of my PDF, pdfauthor = My name, pdfsubject= The subject, pdfkeywords = keyword1 keyword2 keyword3} 
\hypersetup{colorlinks = true, linkcolor = green, anchorcolor = red, citecolor = blue, filecolor = red, pagecolor = red, urlcolor = red}

\begin{document}

   \title{Observational scalings testing modified gravity
}

 \volnopage{ {\bf 20XX} Vol.\ {\bf X} No. {\bf XX}, 000--000}
   \setcounter{page}{1}

   \author{A. Amekhyan\inst{1}, S. Sargsyan\inst{1}, A. Stepanian\inst{1}
   }
%% Here is an example of three authors come from different institutes.
%% For single author or all the authors from an institute, use "\inst{}" only

   \institute{ Center for Cosmology and Astrophysics, Alikhanian National Laboratory, Yerevan, Armenia; {\it arman.stepanian@yerphi.am}\\
%% Please give the E-mail address of the author, to whom future correspondence and
%% offprint requests will be sent.
\vs \no
   {\small Received 20XX Month Day; accepted 20XX Month Day}
}

\abstract{We consider different observational effects to test modified gravity approach involving the cosmological constant in the common description of the dark matter and the dark energy. We obtain upper limits for the cosmological constant by studying the scaling relations for 12 nearby galaxy clusters, the radiated power from gravitational waves and the Tully-Fisher relation for super spiral galaxies. Our estimations reveal that, for all these cases the upper limits for $\Lambda$ are consistent with its actual value predicted by the cosmological observations. 
}

   \authorrunning{A. Amekhyan, S. Sargsyan, A. Stepanian}            %author_head in even pages
   \titlerunning{Observational scalings testing modified gravity}  % title_head in odd pages
   \maketitle

%________________________________________________ sections below
% 
\section{Introduction}
\label{}

A new perspective for describing the dark sector - the dark matter and dark energy - is provided by the modification of gravity based on Newton theorem (\cite{G,GS1,GS2}). Namely, within that approach, the weak-field modification of General Relativity (GR) for spherically symmetric case is written as
\begin {equation} \label{mod}
g_{00} = 1 - \frac{2 G m}{c^2r} - \frac{\Lambda r^2}{3}\,; 
\qquad g_{rr} = \left(1 - \frac{2 G m}{c^2r} - \frac {\Lambda r^2}{3}\right)^{-1}\ ,
\end {equation} 
where the cosmological constant $\Lambda$, as a fundamental constant (\cite{GS4}) is entered self-consistently in the gravity equations. In this sense, this metric can explain the accelerated expansion of the Universe and the dynamics of DM in astrophysical configurations (\cite{GS1}) simultaneously without any further free parameter.

The above metric is obtained by considering the most general function for force $\mathbf{F}(r)$ satisfying Newton's theorem which is equal to (\cite{G,GS1,G1}) 
\begin{equation}\label{FandU}
\mathbf{F}(r) = \left(-\frac{A}{r^2} + Br\right)\hat{\mathbf{r}}\ .
\end{equation}

Importantly, within the McCrea-Milne cosmology the constant $B$ in the second term in Eq.(\ref{FandU}) is corresponding to the cosmological constant  $\Lambda$ (\cite{G,GS1}).  Then, the sign of the cosmological constant $\Lambda$ corresponds to vacuum solutions for General Relativity equations and their isometry groups as shown in Table \ref{tab1}.

\begin{table} 
\centering
\caption{The background geometries of GR according to the sign of $\Lambda$}
{
\begin{tabular}{ |p{0.96cm}||p{2.8cm}|p{2.1cm}|p{1.5cm}|}
\hline
\multicolumn{4}{|c|}{Background geometries} \\
\hline
Sign& Spacetime&Isometry group&Curvature\\
\hline
$\Lambda > 0$ &de Sitter (dS) &O(1,4)&+\\
$ \Lambda = 0$ & Minkowski (M) & IO(1,3)&0 \\
$\Lambda <0 $ &Anti de Sitter (AdS) &O(2,3)&-\\
\hline
\end{tabular}\label{tab1}
}
\end{table}
\noindent 

One of the remarkable features of Eq.(\ref{FandU}) is that, the force, in contrast the pure Newtonian gravity, defines a non-force-free field inside a spherical shell. In fact, this feature agrees with observational indications that galactic halos do determine features of galactic disks (\cite{Kr}). In this sense the weak-field GR will be able to describe the observational features of galactic halos (\cite{G,Ge}), of groups and clusters of galaxies (\cite{GS2}).

The metric in Eq.(\ref{mod}) has been already known as Schwarzschild-de Sitter solution. However, its deduction according to Newton theorem enables one to describe the dynamics of astrophysical structures such as galaxy binaries, groups and clusters in the context of $\Lambda$-gravity (\cite{G,GS2}). The current reported value of $\Lambda$ (\cite{Pl}) is

\begin{equation}\label{pl}
\Lambda = 1.11 \times 10^{-52} \quad m^{-2}
\end{equation}
the previous studies have shown the possibility of the description of galaxy scale effects with metric (\ref{mod}) (\cite{G,Ge,GS2}), we now will aim to test certain observational scalings for galaxies and their clusters, including for the super spirals. The extreme observational samples are known for their efficiency at least for ruling out certain models. We also consider the data of gravitational waves.

\section{Scaling relations}

We use scaling relations from the data of \cite{Desai} for 12 galaxy clusters. Since all of them are considered to be virialized, one can use the virial theorem in the context of metric (\ref{mod}) i.e.
\begin{equation}\label{virL}
\sigma^{2}=\frac{GM}{r} - \frac{\Lambda c^{2}r^{2}}{3},
\end{equation}
where $\sigma$ is the velocity dispersion and $M=\frac{4\pi}{3}r^{3}_{vir}\rho$ is total dynamical mass of cluster, to obtain a possible constraint over the value of $\Lambda$. Thus, in the context of $\Lambda$-gravity, the scaling relations obtained in \cite{Desai} will be written as

\begin{equation}\label{Scale}
\ln\bigg(\frac{\rho_c r_c}{M_{\odot} pc^{-2}}\bigg)=(-0.07^{+0.05}_{-0.06}) \ln\bigg(\frac{M_{200}}{M_\odot}\bigg) +(9.41^{+2.07}_{-1.80}).
\end{equation}
Here, the $\rho_c$ and $r_c$ stand for the central density and core  radius respectively. Meantime, the $M_{200}$ is the total mass enclosed in the radius $r_{200}$ within which the total density
is 200 times denser than the critical density i.e. $H^2 = \frac{8 \pi  G \rho_{crit}}{3}$. Thus, by considering Eq.(\ref{virL}) and Eq.(\ref{Scale}) we will get the following upper limit relation for $\Lambda$
\begin{equation}\label{Lambdalimit}
\Lambda \leq \frac{1}{c^2}  (600 \times H^2)(1- \frac{1}{(\frac{\rho_c  r_c}{(\rho_c + \mathbb{E}(\rho_c)) (r_c + \mathbb{E}(r_c))})^{0.07}}),
\end{equation}
where $\mathbb{E}$ stands for reported error.
The results of calculations are presented in the following Table \ref{tab2}.
\begin{table}
\caption{Constraints obtained for $\Lambda$ by studying the scaling relations}
\centering
\resizebox{\columnwidth}{!}
{
\begin{tabular}{ |p{3.2cm}|p{2.4cm}|p{2.4cm}||p{2.1cm}|| }
\hline
\multicolumn{4}{|c|}{ \textbf{Upper limits for $\Lambda$}} \\ \hline
\hline
Cluster& $\rho_c$ ($10^{-3}M_{\odot} pc^{-3}$)  & $r_c$ (kpc)&$\Lambda$ ($m^{-2}$) $\leq$\\ 
\hline
A133 &   11.68$\substack{ +0.02 \\ -0.02 }$ & 102.01$\substack{+0.08 \\ -0.11 }$ &8.49 $\times$ 10 ${}^{-52}$ \\
\hline
A262 &  5.17$\substack{ +0.87\\ -0.89}$ & 136.36$\substack{+5.40 \\-5.49 }$ &5.68 $\times$ 10 ${}^{-50}$\\
\hline
A383 & 9.63$\substack{+0.62 \\ -0.78}$ & 121.45$\substack{ +3.95\\-4.94 }$ &3.50 $\times$ 10 ${}^{-50}$\\
\hline
A478  & 3.39$\substack{+0.72 \\ -0.84}$ & 286.14$\substack{+30.41 \\-35.62 }$ &5.96 $\times$ 10 ${}^{-50}$\\
\hline
A907 & 4.15$\substack{ +0.42\\ -0.51 }$ & 208.96$\substack{+10.66 \\-12.98 }$ &4.42 $\times$ 10 ${}^{-50}$\\
\hline
A1413 &  6.27$\substack{ +0.49\\ -0.53 }$ & 154.68$\substack{+6.06 \\ -6.61 }$ &3.27 $\times$ 10 ${}^{-50}$\\
\hline
A1795 & 7.15$\substack{ +0.68 \\ -0.79 }$ & 131.89$\substack{+6.32 \\ -7.33}$ &4.07 $\times$ 10 ${}^{-50}$\\
\hline
A1991 & 111.22$\substack{ +0.83\\ -0.92 }$ & 11.15$\substack{+0.04 \\-0.04 }$ &4.20 $\times$ 10 ${}^{-50}$\\ 
\hline
A2029  &  9.39$\substack{ +0.66\\ -0.76 }$ & 134.31$\substack{+4.72 \\-5.45 }$ &3.27 $\times$ 10 ${}^{-50}$\\
\hline
A2390 &  5.83$\substack{ +0.22\\ -0.23 }$ & 137.18$\substack{+2.60 \\ -2.81}$ &1.69 $\times$ 10 ${}^{-50}$\\
\hline
RX J1159+5531 & 41.06$\substack{ +1.33\\-1.19 }$ & 34.07$\substack{+0.55 \\-0.49 }$ &1.12 $\times$ 10 ${}^{-50}$\\
\hline
MKW 4 & 102.4$\substack{ +0.92\\-0.98 }$ & 10.31$\substack{+0.04 \\ -0.04}$&5.06 $\times$ 10 ${}^{-51}$\\
\hline
\end{tabular}\label{tab2}
}
\end{table}
As one can see from the table, the  obtained upper limits of $\Lambda$ for each given cluster is in agreement with its current numerical value (\cite{Pl}).

Here, we want to stress that in general it is possible to study different samples of clusters with different parameters for scaling relations. However, the particular importance of the studied sample is the fact that being a virialized structures we can use the basic fundamental and theoretical relations such as Eq.(\ref{virL}) instead of different empirical models.

\section{Gravitational waves' radiated power}
The existence of gravitational waves (GW) is among the earliest predictions of GR, introduced by Einstein himself (\cite{GW1, GW2}). However, it took one hundred year that first GWs were discovered. In the following, the physics of GWs is discussed in the presence of $\Lambda$. 

Although several astrophysical scenarios can lead to the production of GWs, all the recent detected GWs have been produced by binary systems. Among them, five detections have confirmed that GWs are produced during the merging of two black holes (BH) (\cite{LIGO1, LIGO2,LIGO3,LIGO4,LIGO5}). As one of the observable quantities, we study the radiated power in GW events. As a first step, we have to find the quadrupole moments $\ddot{Q_{ij}}$ which are defined as
\begin{equation}\label{Quadrupole}
\overline{h_{ij}}= \frac{2G}{c^4 r} \ddot{Q_{ij}}(t-\frac{r}{c}) \quad\ i,j=1,2,3
\end{equation}
where $\overline{h_{\mu\nu}} = h_{\mu\nu} - \frac{1}{2} h \eta_{\mu\nu}$ which is called trace-reversed perturbation. The power radiated away during the merging process at null infinity is calculated using the following formula
\begin{equation}\label{Pf}
P= \frac{G}{5 c^5} ((\frac{d ^3 J_{ij}}{dt^3})^2 (\frac{d^3 J^{ij}}{dt^3})^2)
\end{equation}
where $J_{ij}$ is the trace-free part of $\ddot{Q_{ij}}$. Thus, according to \cite{BB}, for two compact objects with mass $M_1$ and $M_2$ orbiting around the circle with radius $r$ with an angular velocity $\omega$, the radiated power $P$, up to first order in $\Lambda$, will be
\begin{equation}\label{Pf}
P= \frac{32 G}{5 c^5} (\frac{M_1 M_2}{M_1 + M_2})^2 a^4 \omega^6 (1+\frac{5 \Lambda c^2}{12 \omega^2}).
\end{equation}

Thus according to \cite{Ligo} for each case, we will have the limits for $\Lambda$ illustrated in Table \ref{tab3}. In this case the obtained upper limits are much larger compared to other measurements which is due to the sensitivity and difficulty of detection of GWs. Namely, it should be recalled that in contrast to other measurements, the first evidence for the existence of  GWs has been observed couple of years ago. Accordingly, the accuracy of reported values as well as the corresponding errors are smaller than those of other experiments and observations. However, it is worth mentioning that despite the current obtained limits for $\Lambda$, the prospective of increase in the accuracy of measurements for detection and analysis of GWs can be regarded as an important and essential test for checking the validity of different modified theories of gravity.

For all of these cases the error limits of $\Lambda$ cover the current observed value. The importance of such analysis lies on the fact that being totally independent from relativistic cosmology, it confirms the validity the current value of $\Lambda$.

\begin{table}
\caption{Constraints obtained for $\Lambda$ by studying the GWs}
\centering
\resizebox{\columnwidth}{!}
{
\begin{tabular}{ |p{1.8cm}|p{2.4cm}|p{1.8cm}||p{2.8cm}|| }\hline
\multicolumn{4}{|c|}{ \textbf{Upper limits for $\Lambda$}} \\ \hline
\hline
GW&  P (erg s${}^{-1}$) & $\omega$ (Hz)& $\Lambda$ ($m^{-2}$) $\leq$\\
\hline
GW 150914&$3.6 ^{+0.5}_{-0.4} \times 10^{56}$ &  75 & $2.08 \times 10^{-14}$\\
\hline
GW 151226&$3.3 ^{+0.8}_{-1.6} \times 10^{56}$& 210 & $2.85 \times 10^{-13}$\\
\hline
GW 170104&$3.1 ^{+0.7}_{-1.3} \times 10^{56}$&  $80-99.5$ & $(3.85-5.96) \times 10^{-14}$\\
\hline
GW 170608&$3.4 ^{+0.5}_{-1.6} \times 10^{56}$&$226.5-305$ & $(2.01-3.64) \times 10^{-13}$\\
\hline
GW 170814&$3.7 ^{+0.5}_{-0.5} \times 10^{56}$& $ 77.5-101.5$ & $(2.16-3.71) \times 10^{-14}$\\

\hline
\end{tabular}\label{tab3}
}
\end{table}

\section{Tully-Fisher relation for super spirals}
In this section, we check Tully-Fisher (TF) relation (\cite{TF}) for recently analyzed group of galaxies called  \textit{super spirals }(SS) (\cite{Ogle}). Indeed, among these galaxies those with stellar mass $M_s > 10^{11.5} M_{\odot}$ show a non-conventional behavior regarding the baryonic TF (BTF) index $b$. Namely, the established BTF index breaks from $3.75 \pm 0.11$ to $0.25 \pm 0.41$ above the rotation velocity of $\approx$ 340 $km/s$. Before starting the analysis, it should be stated that generally objects with some extreme nature/behavior are regarded as useful tools to pose constraints on the different parameters of various modified theories of gravity and even rule them out (\cite{Ext1,Ext2}).

The BTF relation states that
\begin{equation}\label{BTF}
M_{baryonic} \propto V^b_c
\end{equation}
where $V_c$ is the circular velocity of the galaxy. Thus, by replacing Newtonian gravity with $\Lambda$-gravity we will have the following relation
\begin{equation}\label{BTFL}
(\frac{V_c - \mathbb{E}(V_c)}{V_c})^b  \leq 1- \frac{\Lambda c^2 r^3}{3 G M_{baryonic}} 
\end{equation}
in which $\mathbb{E}(V_c)$ is the error limit of circular velocity reported from observations. Considering
above relation, we can obtain the upper limits of $\Lambda$. The results have been illustrated in Table \ref{tab4}. 

\noindent
Meantime, it is also possible to obtain the upper limits of $\Lambda$ by considering the maximum error limit of BTFR index reported for SS galaxies i.e.
\begin{equation}\label{BTFLlimit}
\Lambda \leq (1 - V_c^{\mathbb{E}(b)})\frac{3 G M_{baryonic}}{c^2 r^3} 
\end{equation}
where $\mathbb{E}(b) = \pm 0.41$. Here it is important to mention that, since the $\mathbb{E}(b)$ = 0.41 is the maximum error for all SS galaxies and in principle it could have smaller error for each case, the Eq.(\ref{BTFLlimit}) will give us the absolute upper limit for $\Lambda$ ever possible to obtain based on the SS data. In this case, the obtained limits have been illustrated in Table \ref{tab5}

Considering the obtained upper limits of $\Lambda$ based on both analyses, we can conclude that they are fully consistent with the cosmological observations (\cite{Pl}). 

\begin{table}
\caption{Constraints obtained for $\Lambda$ by studying the super spiral galaxies}
\centering
\resizebox{\columnwidth}{!}
{
\begin{tabular}{ |p{4.4cm}|p{2.4cm}|p{2.1cm}|p{1.2cm}||p{2cm}|| }\hline
\multicolumn{5}{|c|}{ \textbf{Upper limits for $\Lambda$ by considering the error limits of velocity}} \\ \hline
\hline

Galaxy & $\log M_\mathrm{stars} (M{}_{\odot})$ & $\log M_\mathrm{gas} (M{}_{\odot})$ & $r$ (Kpc) & $\Lambda$ $(m^{-2})$ $\leq $ \\ \hline
2MASX J09394584$+$0845033&   11.45 & 10.7 & 14 & 1.85 $\times 10^{-48}$ \\ \hline
SDSS J095727.02$+$083501.7&  11.60 & 10.4 & 31 & 1.84 $\times 10^{-50}$ \\ \hline
2MASX J10222648$+$0911396&   11.42 & 10.5 & 33 & 1.71 $\times 10^{-49}$ \\ \hline
2MASX J10304263$+$0418219&   11.66 & 10.7 & 30 & 2.54 $\times 10^{-50}$ \\ \hline
2MASX J11052843$+$0736413&   11.59 & 10.8 & 54 & 3.08 $\times 10^{-51}$ \\ \hline
2MASX J11232039$+$0018029&   11.43 & 10.6 & 45 & 3.28 $\times 10^{-51}$ \\ \hline
2MASX J11483552$+$0325268&   11.42 & 10.5 & 31 & 1.98 $\times 10^{-49}$ \\ \hline
2MASX J11535621$+$4923562&   11.64 & 10.8 & 19 & 1.42 $\times 10^{-48}$ \\ \hline
2MASX J12422564$+$0056492&   11.24 & 10.2 & 14 & 1.34 $\times 10^{-48}$ \\ \hline
2MASX J12592630$-$0146580&   11.23 & 10.2 & 20 & 3.98 $\times 10^{-49}$ \\ \hline
2MASX J13033075$-$0214004&   11.37 & $<$ 10.4 & 23 & 3.05 $\times 10^{-49}$ \\ \hline
SDSS J143447.86$+$020228.6&  11.60 & 10.7 & 26 & 1.70 $\times 10^{-50}$ \\ \hline
2MASX J15154614$+$0235564&   11.74 & 10.7 & 41 & 9.42 $\times 10^{-51}$ \\ \hline
2MASX J15404057$-$0009331&   11.39 & 10.4 & 30 & 1.45 $\times 10^{-49}$ \\ \hline
2MASX J16014061$+$2718161&   11.63 & 10.6 & 29 & 4.10 $\times 10^{-50}$ \\ \hline
2MASX J16184003$+$0034367&   11.67 & 10.6 & 40 & 1.27 $\times 10^{-50}$ \\ \hline
2MASX J16394598$+$4609058&   11.74 & 10.9 & 31 & 1.61 $\times 10^{-49}$ \\ \hline
2MASX J20541957$-$0055204&   11.41 & 10.6 & 37 & 1.29 $\times 10^{-49}$ \\ \hline
2MASX J21362206$+$0056519&   11.47 & 10.4 & 29 & 2.34 $\times 10^{-49}$ \\ \hline
2MASX J21384311$-$0052162&   11.20 & 9.9 & 15 & 1.06 $\times 10^{-48}$ \\ \hline
2MASX J21431882$-$0820164&   11.13 & 9.9 & 18 & 1.06 $\times 10^{-49}$ \\ \hline
2MASX J22073122$-$0729223&   11.20 & 10.3 & 31 & 1.24 $\times 10^{-50}$ \\ \hline
2MASX J23130513$-$0033477&   11.20 & 10.3 & 19 & 3.91 $\times 10^{-49}$ \\ \hline

\end{tabular}\label{tab4}
}
\end{table}

\begin{table}
\caption{Constraints obtained for $\Lambda$ by studying the super spiral galaxies}
\centering
\resizebox{\columnwidth}{!}
{
\begin{tabular}{ |p{4.4cm}|p{2.4cm}|p{2.1cm}|p{1.2cm}||p{2cm}|| }\hline
\multicolumn{5}{|c|}{ \textbf{Upper limits for $\Lambda$ by considering the error limits of BTFR index}} \\ \hline
\hline

Galaxy & $\log M_\mathrm{stars} (M{}_{\odot})$ & $\log M_\mathrm{gas} (M{}_{\odot})$ & $r$ (Kpc) & $\Lambda$ $(m^{-2})$ $\leq $ \\ \hline
2MASX J09394584$+$0845033&   11.45 & 10.7 & 14 & 1.48 $\times 10^{-51}$ \\ \hline
SDSS J095727.02$+$083501.7&  11.60 & 10.4 & 31 & 7.25 $\times 10^{-52}$ \\ \hline
2MASX J10222648$+$0911396&   11.42 & 10.5 & 33 & 1.46 $\times 10^{-51}$ \\ \hline
2MASX J10304263$+$0418219&   11.66 & 10.7 & 30 & 1.90 $\times 10^{-51}$ \\ \hline
2MASX J11052843$+$0736413&   11.59 & 10.8 & 54 & 6.76 $\times 10^{-52}$ \\ \hline
2MASX J11232039$+$0018029&   11.43 & 10.6 & 45 & 5.60 $\times 10^{-52}$ \\ \hline
2MASX J11483552$+$0325268&   11.42 & 10.5 & 31 & 1.29 $\times 10^{-51}$ \\ \hline
2MASX J11535621$+$4923562&   11.64 & 10.8 & 19 & 2.63 $\times 10^{-51}$ \\ \hline
2MASX J12422564$+$0056492&   11.24 & 10.2 & 14 & 1.30 $\times 10^{-51}$ \\ \hline
2MASX J12592630$-$0146580&   11.23 & 10.2 & 20 & 8.63 $\times 10^{-52}$ \\ \hline
2MASX J13033075$-$0214004&   11.37 & $<$ 10.4 & 23 & 1.32 $\times 10^{-51}$ \\ \hline
SDSS J143447.86$+$020228.6&  11.60 & 10.7 & 26 & 1.64 $\times 10^{-51}$ \\ \hline
2MASX J15154614$+$0235564&   11.74 & 10.7 & 41 & 4.91 $\times 10^{-52}$ \\ \hline
2MASX J15404057$-$0009331&   11.39 & 10.4 & 30 & 1.44 $\times 10^{-51}$ \\ \hline
2MASX J16014061$+$2718161&   11.63 & 10.6 & 29 & 7.52 $\times 10^{-52}$ \\ \hline
2MASX J16184003$+$0034367&   11.67 & 10.6 & 40 & 1.34 $\times 10^{-51}$ \\ \hline
2MASX J16394598$+$4609058&   11.74 & 10.9 & 31 & 8.37 $\times 10^{-52}$ \\ \hline
2MASX J20541957$-$0055204&   11.41 & 10.6 & 37 & 1.39 $\times 10^{-51}$ \\ \hline
2MASX J21362206$+$0056519&   11.47 & 10.4 & 29 & 1.26 $\times 10^{-51}$ \\ \hline
2MASX J21384311$-$0052162&   11.20 & 9.9 & 15 & 9.29 $\times 10^{-52}$ \\ \hline
2MASX J21431882$-$0820164&   11.13 & 9.9 & 18 & 6.70 $\times 10^{-52}$ \\ \hline
2MASX J22073122$-$0729223&   11.20 & 10.3 & 31 & 1.86 $\times 10^{-51}$ \\ \hline
2MASX J23130513$-$0033477&   11.20 & 10.3 & 19 & 1.07 $\times 10^{-51}$ \\ \hline

\end{tabular}\label{tab5}
}
\end{table}

\section{Conclusions}
We have studied the compatibility of modified gravity i.e. weak-field General Relativity with a cosmological constant, by considering several types of observational data, i.e. the scaling relations of galaxy clusters, the radiated power from GWs and the BTF relation.  This paper continues previous analyses in which, by considering both relativistic and non-relativistic effects, various upper limits have been reported for the cosmological constant $\Lambda$ (\cite{M,J1,J2,L,SK, BH}). 

Since the formalism of $\Lambda$-gravity enables us to study both relativistic and non-relativistic effects, we have considered different tests corresponding to each of them.
First, we have obtained constraints for $\Lambda$ by analyzing the data of 12 galaxy clusters reported by Chandra X-ray satellite. In this case, although the upper limits are very close to the current reported value of $\Lambda$, they mark no inconsistency with cosmological observations.

Next, we have calculated the radiated power from GWs and obtained the limits of $\Lambda$ accordingly. Although the obtained limits are less tight than the previous test, for all of the analyzed cases, the current reported value of $\Lambda$ is smaller than the obtained upper limits and hence they fit the predictions of the considered modified gravity. 

Finally, we have checked the recently analyzed super spiral galaxies. It should be stressed that due to their non-typical properties, the super spirals are considered as one of the efficient tests to check the validity of different theories of gravity, as shown also by our analysis. 
  
%\bibliographystyle{raa}
%\bibliography{bibtex}

\end{document}